\documentclass[useAMS]{mn2e}
 \usepackage{times}
 \usepackage{epsfig}
 \usepackage{amssymb}

 \title[Colliding with G2]
       {Colliding with G2 near the Galactic Centre: a geometrical approach}

 \author[R. de la Fuente Marcos and C. de la Fuente Marcos]
        {R.~de~la~Fuente~Marcos
         and
         C. de la Fuente Marcos\thanks{E-mail: nbplanet@fis.ucm.es} \\
         Universidad Complutense de Madrid,
         Ciudad Universitaria, E-28040 Madrid, Spain}
 \date{Accepted 2013 June 20.
       Received 2013 June 19;
       in original form 2013 June 5}
 \pubyear{2013}
 
\begin{document}
  \maketitle

  \begin{abstract}
     The object G2 will pass within $\sim$100 au from Sgr~A$^{*}$ in 2014. 
     Due to its very short periapse, the study of the dynamical evolution 
     of this object in the short-term future may offer some insight into 
     the region surrounding the supermassive black hole at the centre of 
     the Galaxy. With this scenario in mind, it has recently been proposed 
     by Bartos et al. that, prior to its perinigricon, G2 will likely 
     experience multiple encounters with members of the black hole and 
     neutron-star populations believed to orbit near the Galactic Centre. 
     Here, we further explore this possibility and study the general case 
     for collisions with the G2 object using the latest orbital solutions 
     provided by Phifer et al. and Gillessen et al., and a Monte Carlo 
     approach to estimate the minimum orbit intersection distance (MOID) 
     with G2 as a function of the orbital parameters of the incoming body. 
     Our results indicate that encounters at distances closer than 100 au 
     started to become statistically significant only during the last few 
     years or so. MOIDs under 100 au are statistically more probable for 
     the most dynamically cold orbits. If there is a population of objects 
     moving in low-inclination, low-eccentricity orbits around the central 
     black hole, the highest probability for a close encounter with G2 is 
     found to be in the period 2014 January-March but enhanced activity 
     due to encounters may start as early as 2013 July-August.  
  \end{abstract}


  \section{Introduction}
     G2 is an extremely red source with spatially resolved infrared emission moving on a highly eccentric orbit around the supermassive 
     black hole (SMBH) at the centre of the Galaxy; it was first reported by Gillessen et al. (2012) but it appears in images obtained by 
     the Very Large Telescope (VLT) programmes and others since the early 2000s. The object was initially interpreted as a small gas cloud 
     of nearly 100 au in diameter and mass about three times that of the Earth (Gillessen et al. 2012, 2013a). With a predicted perinigricon 
     of nearly 100 au, such a cloud is not expected to survive its close encounter with the SMBH in 2014. If G2 is indeed a gas cloud and is 
     actually being accreted on to the SMBH, this can be used to gain some knowledge on the immediate neighbourhood of the SMBH (e.g. Anninos 
     et al. 2012; Mo\'scibrodzka et al. 2012). However, the pure gas cloud nature of G2 was challenged soon after discovery (Miralda-Escud\'e 
     2012; Murray-Clay \& Loeb 2012; Scoville \& Burkert 2013) and the latest data obtained with the Keck telescopes (Phifer et al. 2013) 
     suggest that G2 has a gaseous component which is tidally interacting with the central SMBH, but there is also a central star hidden in 
     the nebulosity, perhaps a low-mass T Tauri; the T Tauri scenario was already suggested by Scoville \& Burkert (2013). Alternatively, G2 
     may be the result of a recent merger between the two surviving components of a former triple system that was terminated after a three 
     body exchange involving the central SMBH, resulting in the very high orbital eccentricity observed (Phifer et al. 2013); in this 
     scenario, the star is more massive but its light would be dimmed by surrounding dust. In any case, the presence of a central star would 
     provide the self-gravity required to keep the object compact. If it is a star surrounded by gas, the central SMBH could accrete some of 
     the gas, but moving at several thousand km s$^{-1}$, the star may have just enough inertia to avoid being captured and swallowed up, 
     surviving the periapse passage (Phifer et al. 2013). New VLT observations show that the ionized gas component is currently stretched 
     over more than 1200 au (Gillessen et al. 2013b). G2 has attracted an unusual amount of attention which has translated into additional 
     studies providing conflicting interpretations and more or less detailed predictions on what is expected to happen prior to and during 
     its perinigricon (Burkert et al. 2012; Meyer \& Meyer-Hofmeister 2012; Morris 2012; Narayan, \"Ozel \& Sironi 2012; Schartmann et al. 
     2012; Ballone et al. 2013; Czerny et al. 2013; Sadowski et al. 2013; Saitoh et al. 2013; Yusef-Zadeh \& Wardle 2013).
     
     In a recent paper, Bartos et al. (2013) suggest that if the G2 object is indeed a gas cloud, (ordinary or medium-size) black holes 
     orbiting around the SMBH will spin and heat the gas up to millions of degrees, triggering X-ray emissions. These emissions could be 
     detected by X-ray space telescopes such as {\it Chandra} or {\it NuSTAR}. In their paper, Bartos et al. (2013) predict about 16 
     interactions between G2 and the orbiting black holes. However, these authors did not take into account the actual distribution of the 
     orbital elements of the putative black holes. In fact, the layout of the orbits has a major impact on the outcome of their analysis as 
     the relative velocity can become as high as 25\,000 km s$^{-1}$ under certain circumstances. Using the latest orbital solutions for G2
     obtained by Phifer et al. (2013) and Gillessen et al. (2013b), we approach the topic of possible encounters with G2 making no a priori 
     assumptions on the distribution of the orbital elements of the passing objects, degenerate or otherwise. The results of such study may 
     be of help in understanding what to expect of G2 during the next months. This Letter is organized as follows. In Section 2, we outline 
     our Monte Carlo scheme to compute the minimum orbit intersection distance (MOID) of G2 as a function of the orbital parameters of the 
     incoming bodies. The future event timetable according to the various orbital solutions for G2 is discussed in Section 3. Some important 
     points regarding the detectability in X-rays of these interactions are discussed in Section 4. Section 5 summarizes our conclusions. 

  \section{Mapping G2's MOID}
     We are interested in finding out under what conditions a close encounter between an object moving in an arbitrary Keplerian elliptical 
     orbit and G2 is possible. In our analysis, we will neglect relativistic effects, resulting from the theory of general relativity and 
     gravitational focusing. Assuming purely Keplerian orbits around the central black hole is standard in G2 studies (Phifer et al. 2013). 
     Following Innanen et al. (1972), for two objects separated by a distance larger than three times the outer radius of each of them, the 
     amount of mutual disruption is rather negligible. Therefore, and if we assume a diameter of 100 au for G2, the critical distance 
     between G2 and the incoming object is 300 au; we will consider that the amount of tidal disruption on G2 is negligible if the distance 
     of closest approach is $>$ 300 au. In order to experience a close encounter with the MOID smaller than 300 au, the two orbits must 
     nearly intersect and both objects must be almost in the same spot at the same time. This problem is far from trivial (see e.g. 
     Kholshevnikov \& Vassiliev 1999; Gronchi \& Valsecchi 2013) and is well suited for a brute-force Monte Carlo approach in which the two 
     orbits are extensively sampled in phase space, and the distance between any two points on the orbits is computed until the minimal 
     Euclidean distance is eventually found. In our case, we use the two-body problem expressions in Murray \& Dermott (1999) to generate 
     the orbits. The semimajor axes, $a$, of the orbits of the incoming objects are assumed to be $<$ 1 pc, the eccentricities, $e$, are in 
     the range 0--1, the inclinations, $i$, $\in$ (0, 180)$^{o}$ and both the longitudes of the ascending node, $\Omega$, and the arguments 
     of periapse, $\omega$, $\in$ (0, 360)$^{o}$. If the perinigricon epoch for G2 is known, its true anomaly, $f$, can be mapped into an 
     instant in time and the actual timing of the encounters can be studied in detail. This approach is very advantageous as it makes no a 
     priori assumptions on the characteristics of the orbits of the incoming objects; it also allows the calculation of the relative 
     velocity, $\Delta v$, at the time of the encounter. Our approach is computer intensive but it produces an optimal solution for the 
     minimum distance and allows detailed predictions with very little initial information. With this Monte Carlo scheme, we compute the 
     MOID with respect to G2 for billions of test orbits using a sampling resolution of a few million points per orbit. 
%
%
     \begin{figure}
        \centerline{\hbox{
        \includegraphics[width=0.5\linewidth]{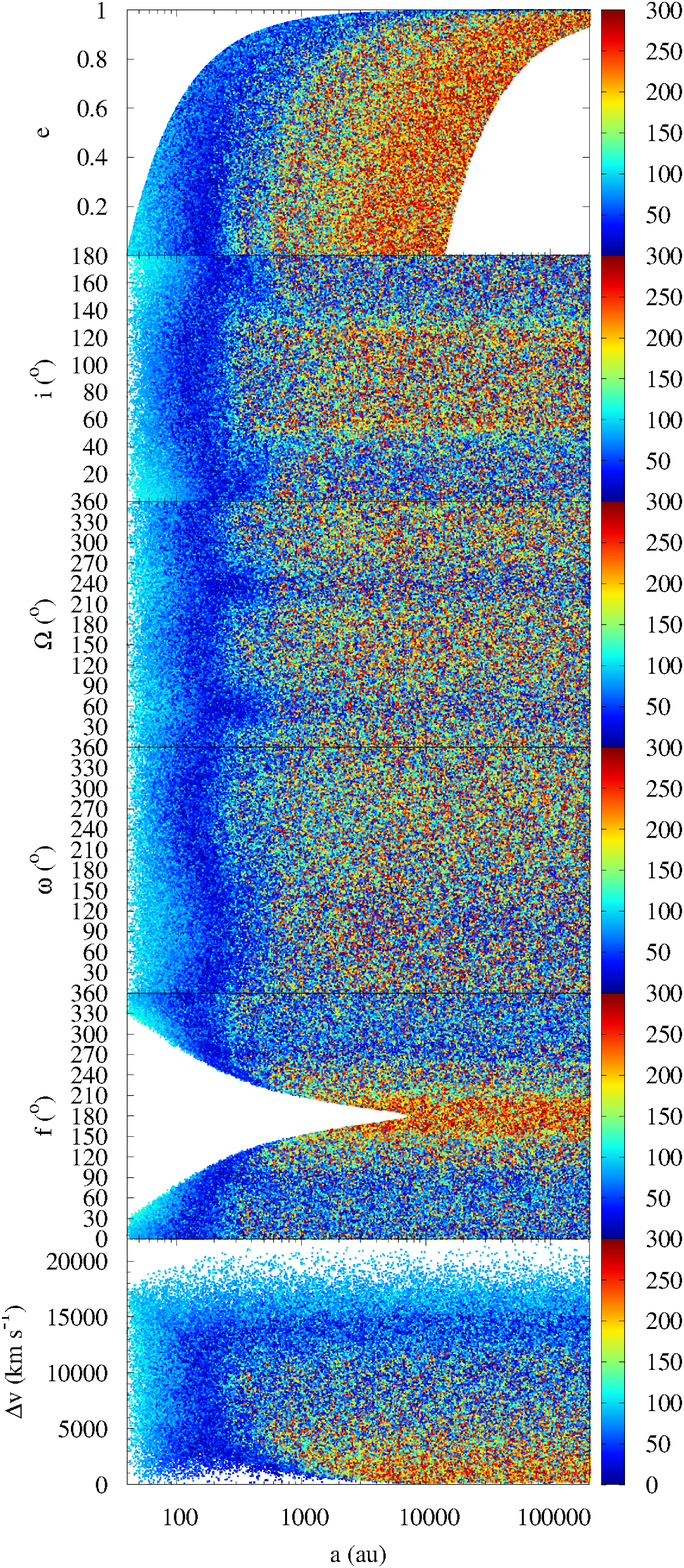}
        \includegraphics[width=0.5\linewidth]{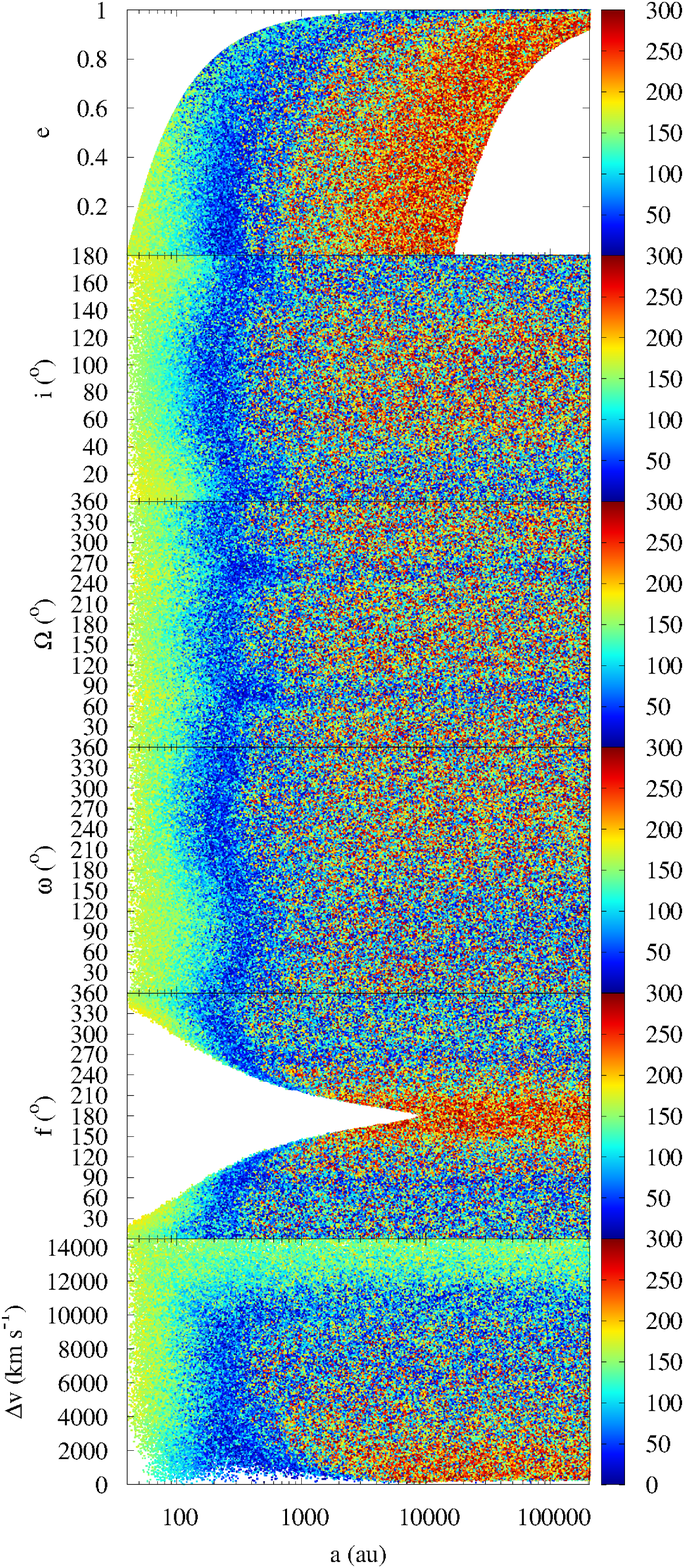}
          }}
        \caption{MOID as a function of the various orbital elements which characterize the orbit of the incoming object: semimajor axis, 
                 $a$, eccentricity, $e$, inclination, $i$, longitude of the ascending node, $\Omega$, and argument of periapse, $\omega$.
                 The value of the MOID as a function of the true anomaly, $f$, of G2 and the relative velocity $\Delta v$ at the time of the
                 close encounter are also shown. Only orbits with periapse $>$ 40 au are displayed. The value of the MOID in au is colour 
                 coded following the scale printed on the associated colour box. The left-hand panels correspond to the nominal orbit of G2
                 in Phifer et al. (2013); the right-hand panels assume the Brackett-$\gamma$ nominal orbit in Gillessen et al. (2013b).}
        \label{map}
     \end{figure}
%
%

\begin{figure*}
        \centering
        \includegraphics[width=\linewidth]{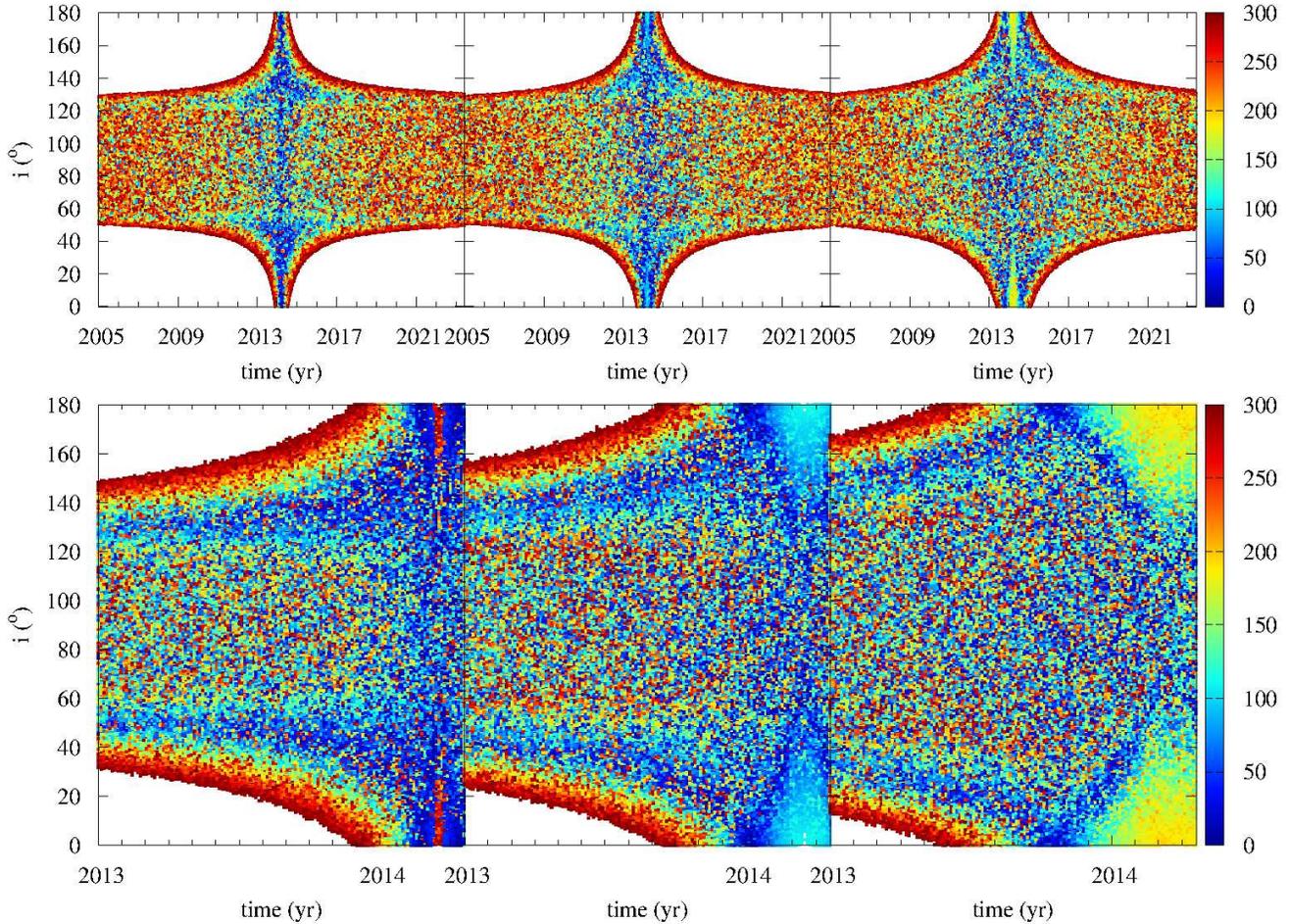}
        \caption{MOID as a function of the inclination of the orbit of the incoming object and the time. The top panels cover the time 
                 interval 2005.00-2023.42, the lower panels show the shorter interval 2013.00-2014.3. Only orbits with periapse $>$ 40 au 
                 are shown. The left-hand panels are for the orbital solution of G2 with $q$ = 62.5 au, the middle panels are for the 
                 nominal orbit in Phifer et al. (2013) and the right-hand panels are for $q$ = 215.4 au. The value of the MOID in au is 
                 colour coded following the scale in the associated colour box (see the text for details).}
        \label{time}
     \end{figure*}

     Our results for the MOID as a function of the various orbital parameters of the trajectory of the incoming object are displayed in Fig. 
     \ref{map} that also includes the true anomaly of G2 and the relative velocity at the time of the close encounter. In this and 
     subsequent figures, the value of the MOID in au is colour coded following the scale printed on the associated colour box. These results 
     are for (left-hand panels) the nominal orbit in Phifer et al. (2013, table 2) and (right-hand panels) the Brackett-$\gamma$ nominal 
     orbit in Gillessen et al. (2013b, table 1). Only the orbits with perinigricon ($q = a (1 - e)$) $>$ 40 au (the effective size of the 
     SMBH; Ghez et al. 2005) are displayed. Similar results are obtained if the error estimates in Phifer et al. (2013) or Gillessen et al. 
     (2013b) are used to construct additional, compatible, orbital solutions. Mapping the true anomaly into time gives us additional 
     information on the actual contribution of the various orbits to possible encounters with G2 in the near future. This time-resolved 
     information for the nominal orbit in Phifer et al. (2013) is shown in Fig. \ref{time} and it clearly indicates that objects moving in 
     low inclination orbits ($i < 20^{o}$, prograde or retrograde) are more likely to experience close encounters with G2. Results in Fig. 
     \ref{time} correspond to three different orbital solutions, one with maximum encounter rate, another one with minimum encounter rate 
     and the third one being intermediate. The nominal orbit ($a$ = 6989 au, $e$ = 0.9814 and $i$ = 121$^{o}$) given by Phifer et al. (2013), 
     middle panels, produces intermediate results. Their orbit with the lowest possible periapse (minimum $a$ and maximum $e$) and lowest 
     inclination ($a$ = 4960 au, $e$ = 0.9874 and $i$ = 124$^{o}$) gives the highest number of close encounters (left-hand panels). Finally, 
     the minimum encounter rate is found for the orbit with the largest perinigricon (maximum $a$ and minimum $e$) and the highest 
     inclination ($a$ = 8756 au, $e$ = 0.9754 and $i$ = 118$^{o}$, right-hand panels). The behaviour is similar for the nominal orbit ($a$ = 
     8324 au, $e$ = 0.9762 and $i$ = 118$^{o}$) given by Gillessen et al. (2013b). As expected, the timing, frequency and strength of the 
     encounters depend significantly on the orbital solution considered for G2. A shorter periapse delays and shortens the most dynamically 
     active phase. In any case, our results indicate that encounters at distances closer than 100 au only started to become statistically 
     significant during the last few years.
%
%
%
%

  \section{A timetable to disaster}
\begin{figure}
       \centering
        \includegraphics[width=\linewidth]{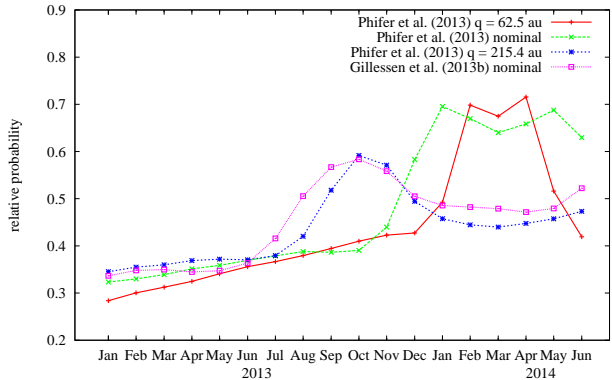}
        \caption{Relative probability of having a close encounter with MOID $<$ 100 au for the nominal orbits in Phifer et al. (2013) and
                 Gillessen et al. (2013b), and also for those with $q$ = 62.5 au and $q$ = 215.4 au (see the text for details).}
        \label{prob}
     \end{figure}

     The object G2 will pass within $\sim$100 au from Sgr~A$^{*}$ in 2014; the time of closest approach is predicted to be 2014.21$\pm$0.14
     or about 2014 mid-March (Phifer et al. 2013). If it is a gas cloud, it will be tidally disrupted during its periapse passage and it is 
     not expected to survive. Therefore, if close encounters with other objects, likely black holes (ordinary or medium size) and neutron 
     stars, are bound to happen prior to its demise, what is the expected timetable of events according to our modelling, i.e., when is the 
     received X-ray emission expected to be more intense? In order to answer this important question, we have computed the probability of G2
     undergoing an encounter at a distance closer than 100 au relative to our sample of encounters with MOID $<$ 300 au for the months prior
     and around the perinigricon epoch. This probability has been calculated as the ratio of events with MOID $<$ 100 au to those with MOID 
     $<$ 300 au. We have obtained this relative probability for the nominal orbits in Phifer et al. (2013) and Gillessen et al. (2013b), and 
     also for the limiting cases, based on the error estimates in Phifer et al. (2013), of $q$ = 62.5 au and $q$ = 215.4 au. These limiting 
     cases have been described in detail above. Our predictions are summarized in Fig. \ref{prob}. Penetrating encounters, those with MOID 
     $<$ 100 au, are dominant if the relative probability is $>$ 0.5. If there is a population of objects surrounding the central SMBH, 
     their observational effects on G2 have markedly different timings and strengths, depending on the adopted orbital solution for G2. The 
     effects are expected to be more important for the shortest perinigricon but start sooner for the wider periapse. If $q$ = 215.4 au and 
     assuming that the current level of emissions is zero, the highest probability of close encounters (and, possibly, the X-ray emissions 
     peak) is expected in 2013 October-November, the nominal model predicts a peak in 2014 January-May and for $q$ = 62.5 au, the maximum is 
     expected in 2014 February-April. In general, the solution obtained by Gillessen et al. (2013b) predicts earlier activity. As pointed 
     out above, not all the passing objects are expected to equally contribute to this probability. The largest contribution is coming from 
     objects moving in orbits with relatively low inclinations (see Fig. \ref{time}). If most objects are following orbits perpendicular to 
     G2's orbital plane, the effects of these encounters will likely be negligible. However, if most objects move in prograde or retrograde 
     orbits with an inclination $<$ 20$^{o}$, dramatic changes on G2 may be observed. If encounters do happen, it should be fairly easy to 
     distinguish among the various orbital solutions for G2 by studying G2's light curve. This can also provide valuable information on the 
     distribution of orbital elements of members of the dark population of stellar remnants located towards the Galactic Centre. 
     Unfortunately, ground-based telescopes cannot observe the area around Sgr~A$^{*}$ between October and February as it remains lost in 
     the Sun's glare; the chances of being able to test the predictions of our models are better for space-borne telescopes like 
     {\it Chandra} as they will miss the region of interest only between November and January.

%
%
%
%
%
%
     \begin{figure}
       \centering
        \includegraphics[width=\linewidth]{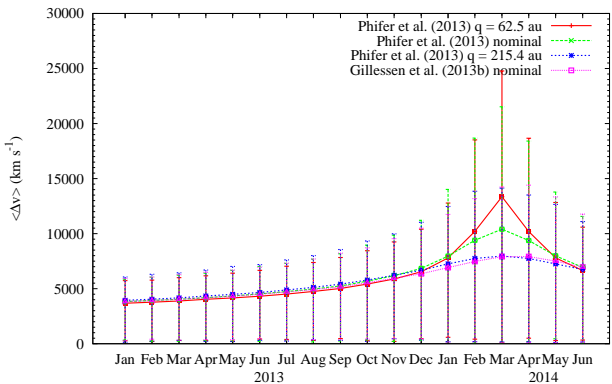}
        \caption{Average relative velocity at the instant of closest approach for close encounters with MOID $<$ 100 au and the four 
                 orbital solutions in Fig. \ref{prob}. The error bars represent the range between the minimum and the maximum relative 
                 velocities not the standard deviation. The distribution is skewed.}
        \label{v}
     \end{figure}
%
%

  \section{Discussion}
     In their paper, Bartos et al. (2013) predict about 16 interactions between G2 and the orbiting black holes during the final 6 months 
     prior to periapse. Our modelling cannot give absolute figures for the number of encounters expected as they strongly depend on the 
     orbits of the objects, the actual distribution of their orbital elements being, at present, unknown. Predictions in Bartos et al. 
     (2013) do not account for the particular details of the orbits of the incoming black holes, so they may be too pessimistic/optimistic 
     depending on the real distribution of the orbital elements of the black holes. If there is a torus/disc-like structure within a few 
     1000 au around Sgr~A$^{*}$ made of degenerate objects and this structure is coplanar to G2 or has low inclination, the observational 
     effects on G2 could be much larger than anticipated. In any case, our calculations indicate that, for the time interval starting on 
     2013 January and ending just before the periapse, the probability of finding MOID $<$ 100 au for G2 is in the range 2--3$\times10^{-3}$ 
     if the orbits of objects moving in the neighbourhood of Sgr~A$^{*}$ are uniformly distributed in orbital parameter space. For an 
     assumed population of 20\,000 objects (black holes or neutron stars; e.g. Morris 1993; Miralda-Escud\'e \& Gould 2000; Freitag, 
     Amaro-Seoane \& Kalogera 2006; O'Leary, Kocsis \& Loeb 2009), the number of encounters is in the range 40--55. The difference with 
     Bartos et al. (2013) can be attributed to their use of a less eccentric orbital solution and also to different assumptions regarding 
     the spatial distribution of incoming objects. In any case, the number of close encounters will be far from negligible and some effects, 
     perhaps dramatic, should be observed before the actual periapse even considering the limited window of observability pointed out above. 
     The objects with the lowest MOIDs with respect to G2 are all moving in highly eccentric orbits ($e >$ 0.6). Encounters at relative 
     velocities $<$ 2000 km s$^{-1}$ tend to be associated to objects with MOIDs $>$ 200 au. Conversely, objects with small MOIDs experience 
     encounters with G2 at much higher relative velocities, up to 21\,500 km s$^{-1}$ for the nominal orbit in Phifer et al. (2013) and as 
     high as 25\,000 km s$^{-1}$ in the most extreme case studied here ($q$ = 62.5 au, see above). Relative velocities for the nominal orbit 
     in Gillessen et al. (2013b) are always lower. The average relative velocity at the instant of closest approach for close encounters 
     with MOID $<$ 100 au and the four orbital solutions discussed above are shown in Fig \ref{v}. MOID $<$ 100 au implies penetrating 
     encounters in which the incoming object, likely a black hole or neutron star, can accrete some gas from G2 and convert a fraction of it 
     into electromagnetic radiation. The error bars in Fig. \ref{v} do not represent the standard deviation, but the actual full range 
     between the minimum and the maximum relative velocities at closest approach for a given epoch. The distribution is skewed. For the 
     nominal orbit in Phifer et al. (2013) and using the mean values of the relative velocities, an encounter lasts over 20 d; this 
     time-scale increases to over 30 d if $q$ = 215.4 au and decreases to 15 d if $q$ = 62.5 au. Encounters last longer for the solutions in 
     Gillessen et al. (2013b). Unfortunately, at these relative speeds and if the mass of the incoming object is less than a few thousand 
     solar masses, the induced X-ray luminosity is most probably several orders of magnitude below the detectability limit of available 
     X-ray observatories. Equation 3 for the X-ray luminosity, $L_x$, in Bartos et al. (2013) applies to black holes and scales as $\Delta 
     v ^{-6}$; even for 1000 M$_{\odot}$ black holes, the range in $L_x$ is 1.0$\times$10$^{28}$--6.4$\times$10$^{29}$ erg s$^{-1}$. 
     Equation 4 in Bartos et al. (2013) applies to neutron stars and scales as $\Delta v ^{-3}$; now, the range in $L_x$ is 
     2.3$\times$10$^{28}$--1.9$\times$10$^{29}$ erg s$^{-1}$. In contrast, the detection limit for observations in the vicinity of the 
     Galactic Centre for the {\it Chandra} X-ray observatory is 10$^{32}$ erg s$^{-1}$. In this context, only multiple simultaneous 
     interactions, incoming black holes with masses $>$ 30\,000 M$_{\odot}$ or (more likely) encounters at low relative velocity (well below 
     the mean value in Fig. \ref{v}) will be able to generate a detectable flux. Although the gas component associated to G2 is now 
     stretched over more than 1200 au and this may increase the effective cross section for encounters, the dimensions of the core appear to 
     be similar to those in 2004 (see fig. 3 in Gillessen et al. 2013b). This suggests that the density of the gas in the stretched regions 
     could be substantially lower than that near the core of the object and the scenario outlined here may not be significantly affected by 
     the stretching. G2 is not only expected to interact with multiple incoming objects but also with relativistic jets from Sgr~A$^{*}$ 
     itself (Yusef-Zadeh \& Wardle 2013). Our calculations suggest that, as far as X-ray emission is concerned, an encounter between the 
     nebular component of G2 and a jet from Sgr~A$^{*}$ --if properly aligned-- could be much easier to detect than the effects of the 
     high-speed encounters studied here unless intermediate-mass black holes are overabundant near the Galactic Centre which is unlikely. 

     The tidal effects on the gas component of G2 following a sequence of close encounters with incoming black holes may also be observable.
     The effect of a runaway SMBH passing through the intergalactic medium has been studied by de la Fuente Marcos \& de la Fuente Marcos 
     (2008) using the impulse approximation. The putative black holes considered here are significantly smaller but their velocities are
     comparable to those in de la Fuente Marcos \& de la Fuente Marcos (2008) and their minimum interaction distance is much smaller (100 au
     versus 50 pc). Their equation 1 gives a value of the velocity impulse induced on G2 $<$ 20 km s$^{-1}$ for an intermediate-mass black 
     hole but it becomes $<$ 1 km s$^{-1}$, the typical value of the sound speed in a gas cloud, for ordinary black holes or neutron stars. 
     This suggests that only medium-sized (or larger) black holes may be able to induce observable changes in the gas component of G2.

  \section{Conclusions}
     Making few a priori assumptions, we have shown that if there is a population of stellar remnants orbiting within the central parsec of 
     the Milky Way, the G2 object should experience a significant number of close, even penetrating encounters, in the months prior to its 
     periapse, perhaps starting as early as 2013 July-August. The timing, frequency and strength of these interactions may provide an 
     improved view of the actual path followed by G2 as well as of the non-luminous or dimmer inhabitants of the immediate neighbourhood of 
     Sgr~A$^{*}$. This makes G2 a potentially powerful tool in the study of the elusive dark component that may roam the core of the Milky 
     Way. In general, our results confirm those in Bartos et al. (2013) although we provide a much more pessimistic evaluation of the 
     outcome of these encounters in terms of G2's X-ray variability as it may be hardly detectable. Our Monte Carlo approach is powerful and 
     general enough to provide a clear view of the general problem of collisions near the Galactic Centre and, in particular, the case of G2.

     Our detailed geometrical and statistical analysis of the theoretically possible close encounters between G2 and other objects moving 
     close to the centre of the Milky Way also provides general clues on the nature of interactions between non-extended objects in that 
     region. Most well studied stars orbiting Sgr~A$^{*}$ move in very eccentric orbits (0.36--0.97), but with shorter periods ($<$ 100 yr, 
     Ghez et al. 2005), Fig. \ref{map} indicates that close encounters are negligible at aponigricon and for inclinations in the range 
     60--90$^{o}$ in both prograde and retrograde orbits. However, and even if close encounters are far less probable under these conditions, 
     their duration doubles or triples with respect to typical encounters closer to Sgr~A$^{*}$; this may translate into long-lasting tidal 
     effects.  

  \section*{Acknowledgements}
     The authors would like to thank the anonymous referee for his/her helpful and quick report. This work was partially supported by the 
     Spanish `Comunidad de Madrid' under grant CAM S2009/ESP-1496. We thank M. J. Fern\'andez-Figueroa, M. Rego Fern\'andez and the 
     Department of Astrophysics of the Universidad Complutense de Madrid (UCM) for providing computing facilities. Most of the calculations 
     and part of the data analysis were completed on the `Servidor Central de C\'alculo' of the UCM and we thank S. Cano Als\'ua for his 
     help during this stage. In preparation of this Letter, we made use of the NASA Astrophysics Data System and the astro-ph e-print server.

\end{document}